\pdfoutput=1
\documentclass[DIV12]{scrartcl}
\usepackage[utf8]{inputenc}
\usepackage[T1]{fontenc}
\usepackage[english]{babel}
\usepackage{authblk}
\usepackage{cite}
\usepackage{graphicx}
\usepackage{amsmath}
\usepackage[squaren,Gray,textstyle,thinqspace]{SIunits}
\setkomafont{disposition}{\normalcolor\rmfamily}

\setcapindent{0mm}
\typearea[current]{current}
\begin{document}
\title{Nonlinear lattice structures based on\\ families of complex nondiffracting beams}
\author{Patrick Rose, Martin Boguslawski, and Cornelia Denz}
\affil{\large Institut für Angewandte Physik and Center for Nonlinear Science (CeNoS), Westfälische Wilhelms-Universität Münster,\\ Corrensstraße~2/4, 48149~Münster, Germany}
\date{}
\maketitle
\begin{abstract}
We present a new concept for the generation of optical lattice waves. For all four families of nondiffracting beams, we are able to realize corresponding nondiffracting intensity patterns in one single setup. The potential of our approach is shown by demonstrating the optical induction of complex photonic discrete, Bessel, Mathieu, and Weber lattices in a nonlinear photorefractive medium. However, our technique itself is very general and can be transferred to optical lattices in other fields like atom optics or cold gases in order to add such complex optical potentials as a new concept to these areas as well.
\end{abstract}
\section*{Introduction}
\label{sec:intro}
\noindent
Due to their technological importance in a multitude of application fields, artificial photonic materials have become a very active research area in recent years. In this context, photonic crystal structures with their unique transmission and reflection spectra provide one of the most influential approaches. In particular, the occurrence of photonic band gaps in these periodically modulated materials~\cite{ref:Yablonovitch1987,ref:John1987} fascinates scientists from all fields and offers novel possibilities to control and guide the propagation of light~\cite{ref:Joannopoulos1997}.

Photonic crystal fibers~\cite{ref:Knight1996} provide a prominent example to demonstrate the relevance of structured materials for photonic applications. Their internal microstructure facilitates tunable zero dispersion wavelengths and very high nonlinearities which make these fibers~-- for instance~-- highly convenient for supercontinuum generation~\cite{ref:Ranka2000}.

While this application illustrates the significant influence of structured materials on light and light propagation, photons in turn can be highly useful to structure material properties as well. The emerging field of laser-assisted material processing, be it for example holographic laser lithography~\cite{ref:Campbell2000} or direct femtosecond laser writing~\cite{ref:Kawata2001}, accounts for this insight.

Combining both aspects~-- light-induced material structuring on the one and the influence of structured media on the propagation characteristics of light on the other hand~-- leads us to the concept of optically induced photonic structures. Materials that change their properties due to light whereupon light itself reacts on the changed material environment provide the necessary link in order to control and guide light by light itself.

However, modulated light fields required for an efficient light induced material structuring also have important applications in other areas. For example optical tweezing and micrometer particle assembly rely on corresponding optical potentials~\cite{ref:Benito2008,ref:Woerdemann2010}, and atom traps for Bose-Einstein condensates are based on optical lattices as well~\cite{ref:Bloch2005}. In this respect, the new concepts for complex photonic lattices presented in the following will add significantly to all these physical fields.
\section*{Nondiffracting beams for optical induction}
\label{sec:optical_induction}
Among all the different realizations of optically induced photonic structures, the optical induction of photonic lattices in photorefractive strontium barium niobate (SBN) crystals~\cite{ref:Efremidis2002,ref:Fleischer2003} is
particularly flexible. This specific approach takes advantage of SBN's electro-optic properties and therewith allows to achieve highly reconfigurable refractive index distributions with various geometries.

Physically, the induced refractive index modulation is caused by the photorefractive effect~\cite{ref:Kukhtarev1979}. Under illumination with a spatially modulated intensity distribution~-- the so-called lattice wave~-- charge carriers in the medium are excited via photoionisation and redistributed in an externally applied electric field. Inside the photorefractive crystal, this redistribution results in a reversible macroscopic space-charge field which in turn modulates the refractive index by means of the electro-optic effect.

This concept of optically induced two-dimensional photonic lattices facilitates new insights in various fields of physical science and has been utilized to demonstrate for instance discrete solitons~\cite{ref:Fleischer2003}, Anderson localization~\cite{ref:Schwartz2007}, as well as Zener tunneling and Bloch oscillations~\cite{ref:Trompeter2006}.

While up to now only rather simple lattice geometries have been studied in this field, we show that the concept of optically structured photonic materials is not limited to such basic patterns. In the following, a new induction approach is introduced which enables the fabrication of photonic structures with various complex geometries.

In order to generate a two-dimensional photonic lattice inside a photosensitive material, the intensity profile of the inducing lattice beam has to be modulated in the two transverse dimensions but invariant in the direction of propagation. All beams satisfying this condition share the property that their transverse spatial frequency components lie on a circle in the corresponding Fourier plane~\cite{ref:Bouchal2003}. Therefore, the general Fourier spectrum of a nondiffracting beam can conveniently be written in polar coordinates as
\begin{equation}
F(\nu, \varphi) = A(\varphi - \varphi_0) \, \delta(\nu - \nu_0) \; ,
\label{eq:fourierspectrum_general}
\end{equation}
with a radial spatial frequency $\nu$ and with $\delta$ being the Dirac delta function, limiting the spectral distribution to a circle with radius $\nu_0$. $A(\varphi)$ gives the complex Fourier spectrum on this circle as a function of the azimuthal angle $\varphi$ while $\varphi_0$ accounts for the possibility to rotate the whole transverse structure by this angle.

This general spectrum indeed leads to ideal nondiffracting beams whose transverse intensity profiles remain invariant for arbitrary propagation distances but at the same time carry infinite energy~\cite{ref:Bouchal2003}. Since actual experimental realizations naturally require a finite total energy of the beam, the nondiffracting character can merely be approximated. Therefore, a rigorous theoretical description of the beam propagation would have to take finite apertures into account, and usually this is modelled as Helmholtz-Gauss beams~\cite{ref:Gori1987,ref:Lopez-Mariscal2006}. The resulting spectrum then is given by a convolution of~(\ref{eq:fourierspectrum_general}) with the spectrum of the transverse envelope. However, we intend to use the given equations for the calculation of the input to our experimental setup and thus refer only to the ideal formulation given above.

Theoretical investigations revealed four different fundamental families of propagation invariant light fields. Depending on the underlying real space coordinate system, a distinction is drawn between discrete, Bessel, Mathieu, and Weber nondiffracting beams~\cite{ref:Bouchal2003,ref:Bandres2004}.

The family of discrete nondiffracting beams can be considered as a set of plane wave interference patterns. The classical two beam interference is a well-known member of this family showing a nondiffracting stripe pattern as its transverse intensity profile. Besides stripes, all other regular plane tilings~-- i.e.\ square, hexagonal, and triangular patterns~-- have been realized with discrete nondiffracting beams as well~\cite{ref:Fleischer2003,ref:Bartal2005,ref:Rose2007}.

While the experimental implementation of discrete nondiffracting beams is comparatively simple, the other three beam families require a far more elaborated approach. By means of computer-controlled light modulators, we are able to spatially modulate both amplitude and phase of an incident light wave. This novel technique permits us to generate arbitrary lattice waves out of all families of nondiffracting beams within one single setup.
\section*{Experimental realization of complex nondiffracting beams}
\label{sec:setup}
\begin{figure}[htb]
\centering
\includegraphics[width=8.4cm]{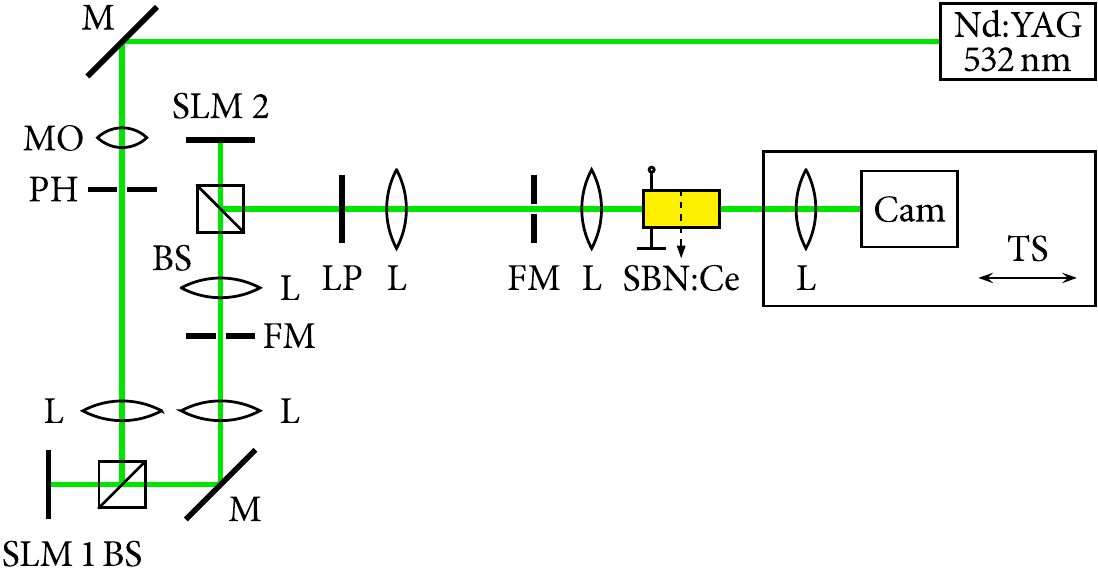}
\caption{Experimental setup for the optical induction of complex nonlinear photonic lattice structures. BS: beam splitter, FM: Fourier mask, L: lens, LP: linear polarizer, M: mirror, MO: microscope objective, PH: pinhole, SLM: spatial light modulator, TS: translation stage.}
\label{fig:setup}
\end{figure}

The experimental setup used for the generation of these complex nondiffracting intensity patterns and their utilization for the optical induction of photonic structures is schematically shown in Figure~\ref{fig:setup}.

A beam from a frequency-doubled Nd:YAG laser at a wavelength of~\unit{532}{\nano \meter} illuminates a first programmable light modulator. This modulator imprints a spatial phase pattern specifically calculated for the desired nondiffracting wave onto the illuminating beam. Subsequently, this first modulator is imaged onto a second one which modulates the amplitude of the beam according to the calculated intensity distribution of the nondiffracting lattice wave. The modulated beam is finally imaged by a high numerical aperture telescope onto the input face of an SBN crystal with a length of \unit{5}{\milli\meter}. Since all nondiffracting beams share the discussed property that their transverse spatial frequency components lie on a circle in the corresponding Fourier plane~\cite{ref:Bouchal2003}, spatial filters within the Fourier planes of the two involved telescopes ensure this constraint.

\begin{figure}[htb]
\centering
\includegraphics{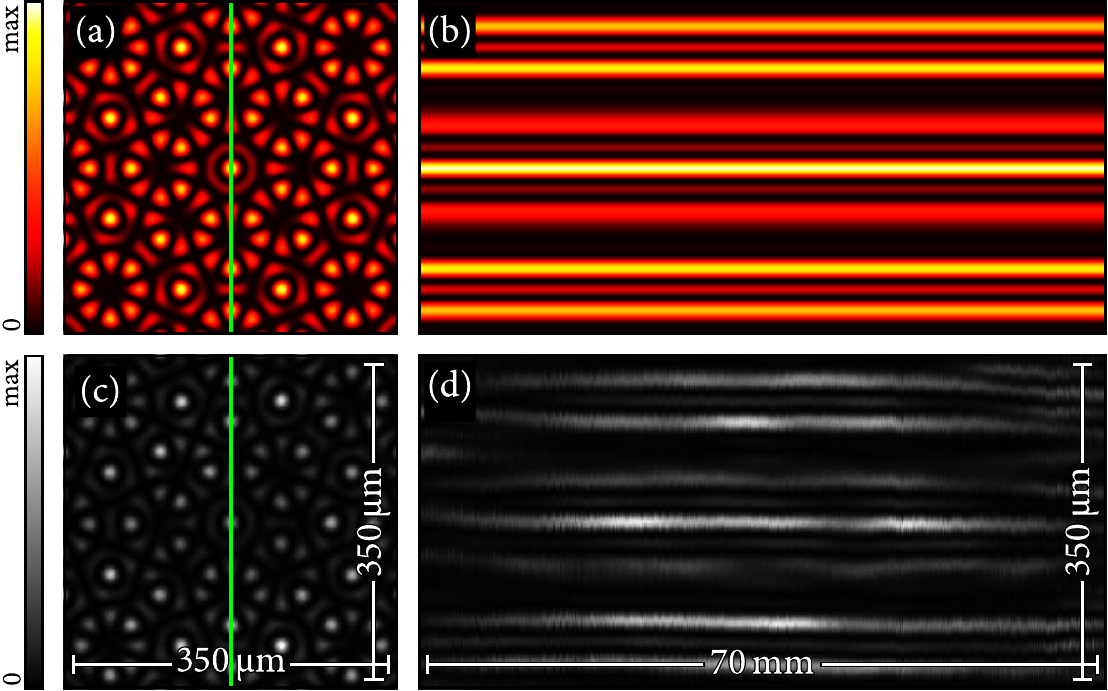}
\caption{Discrete nondiffracting beam with 8-fold rotational symmetry (cf.\ Equation~(\ref{eq:fourierspectrum_discrete_general}), $N = 8$, $m = 0$, $\varphi_0 = 0$). (a)~Calculated and (c)~experimentally observed transverse intensity distribution, (b)~calculated and (d)~experimentally observed nondiffracting intensity profile in longitudinal direction. The green lines in~(a) and~(c) mark the longitudinal planes shown in~(b) and~(d), respectively. All figures are normalized.}
\label{fig:nondiffracting}
\end{figure}

A camera is used to analyze the intensity distribution at the output face of the crystal. In addition, this imaging part of the setup is mounted on a translation stage and can be moved to image different transverse planes of the lattice wave. After removing the crystal, this tool allows for an analysis of the wave's nondiffracting intensity profile in air as well.

To illustrate the nondiffracting propagation of a lattice wave in our setup, we generated a prototypical discrete nondiffracting beam (cf.\ Figure~\ref{fig:nondiffracting}a) and observed its transverse intensity pattern (cf.\ Figure~\ref{fig:nondiffracting}c) at multiple equidistant longitudinal positions. By stacking all these two-dimensional images, we get a three-dimensional intensity distribution of the lattice wave that makes an analysis of the propagation characteristics possible. Figure~\ref{fig:nondiffracting}d shows a longitudinal cross-section of the considered beam and compared to the corresponding perfect discrete nondiffracting wave field with infinite energy (cf.\ Figure~\ref{fig:nondiffracting}b) we see a nondiffracting propagation range of about seven centimeters realized with our setup. This distance considerably exceeds the thickness of all appropriate photosensitive materials.
\section*{Optical induction using discrete nondiffracting beams}
\label{sec:discrete}
\begin{figure}[htb]
\centering
\includegraphics{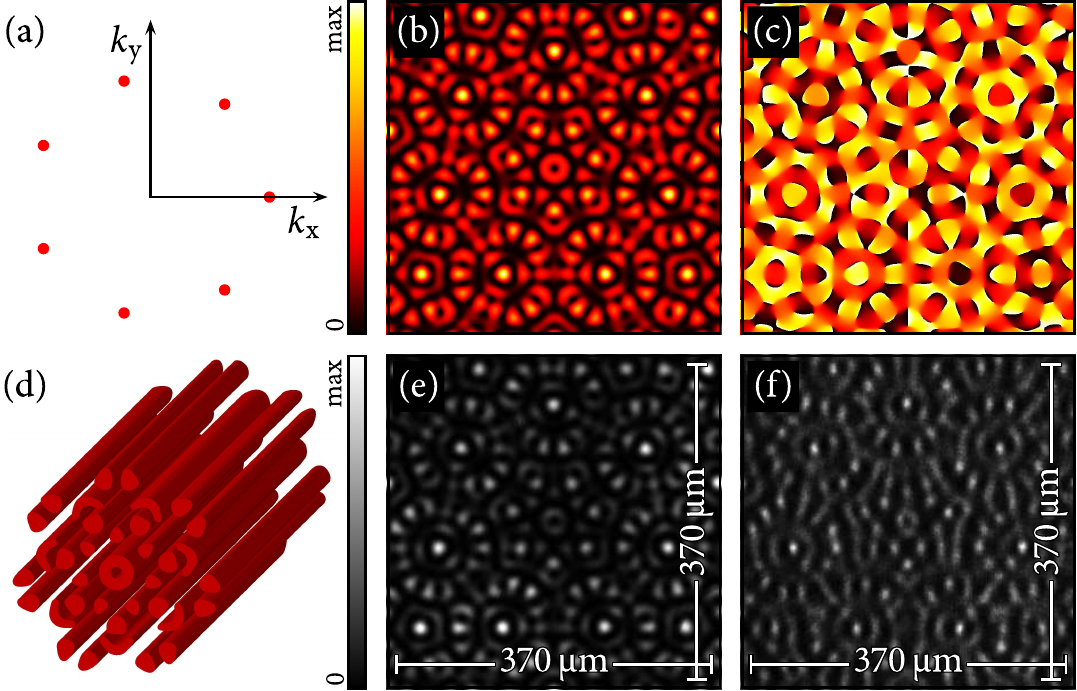}
\caption{Discrete nondiffracting beam ($N = 7$, $m = 1$, $\varphi_0 = 0$). (a)~Calculated transverse intensity distribution in Fourier space and (b)~real space, (c)~corresponding real space phase distribution, (d)~iso-intensity surfaces of the nondiffracting light field at 20\% of the maximum intensity, (e)~experimentally observed transverse intensity profile, (f)~waveguiding in optically induced discrete photonic lattice. Besides~(c) all figures are normalized.}
\label{fig:discrete}
\end{figure}

For typical discrete nondiffracting beams, the azimuthal contribution in Equation~(\ref{eq:fourierspectrum_general}) is given by
\begin{equation}
A(\varphi) = A_0 \exp(\mathrm{i} m \varphi) \sum\limits_{j = 0}^{N - 1} \delta\left(\varphi - \frac{2\pi}{N} j\right)
\label{eq:fourierspectrum_discrete_general}
\end{equation}
representing the Fourier transform of $N$ interfering plane waves with amplitude $A_0$. The resulting Fourier spectrum consists of discrete intensity spots forming the corners of a regular $N$-fold polygon. The phase difference between neighboring spots is determined by the total topological charge $m$.

Besides regular plane tilings, discrete nondiffracting beams can also be used to generate quasicrystallographic patterns~\cite{ref:Freedman2006,ref:Boguslawski2011}, i.e.\ patterns that possess a long-range order but lack the characteristic translational periodicities of crystals (cf.\ Figs.~\ref{fig:nondiffracting}a and~\ref{fig:nondiffracting}c).

The optical induction of such a quasicrystallographic structure~-- exemplarily with 7-fold rotational symmetry in this case~-- is demonstrated in Figure~\ref{fig:discrete}. Based on~(\ref{eq:fourierspectrum_general}) and~(\ref{eq:fourierspectrum_discrete_general}), Figure~\ref{fig:discrete}a depicts the Fourier spectrum of the considered nondiffracting beam ($N = 7$, $m = 1$, $\varphi_0 = 0$).

For the experimental realization, we firstly calculate the transverse real space intensity and phase profiles of the desired beam (cf.\ Figs.~\ref{fig:discrete}b and~\ref{fig:discrete}c, respectively). Then, these numerical cross sections are used as input for our experimental writing setup, and Figure~\ref{fig:discrete}e shows the impressive agreement between the generated lattice wave and the addressed nondiffracting beam (Fig.~\ref{fig:discrete}b). In addition, Figure~\ref{fig:discrete}d depicts iso-intensity surfaces of the simulated nondiffracting beam propagating through the photorefractive crystal.

In order to analyze the actually induced structure, the lattice is illuminated with a broad plane wave, which is then guided by the regions of high refractive index. As a consequence, the modulated intensity distribution at the output of the crystal qualitatively maps the induced refractive index change~\cite{ref:Desyatnikov2006}. Figure~\ref{fig:discrete}f clearly shows the successful induction of the quasicrystalline photonic lattice.
\section*{Photonic Bessel lattices}
\label{sec:bessel}
\begin{figure}[htb]
\centering
\includegraphics{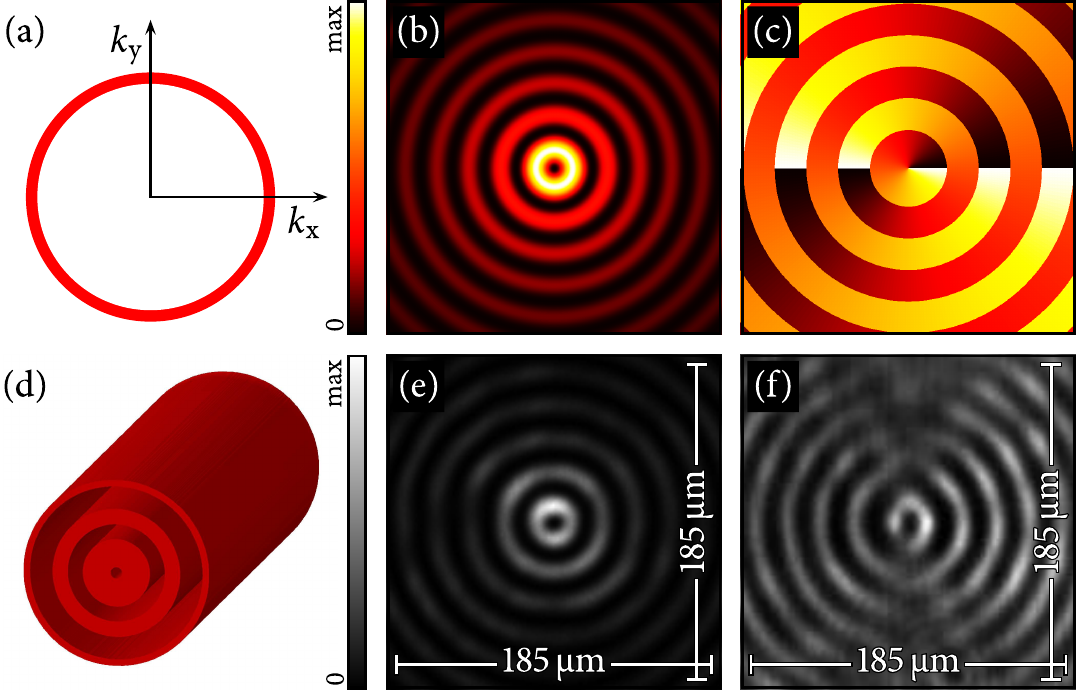}
\caption{First order nondiffracting Bessel beam ($m = 1$). (a)~Calculated transverse intensity distribution in Fourier space and (b)~real space, (c)~corresponding real space phase distribution, (d)~iso-intensity surfaces of the nondiffracting light field at 20\% of the maximum intensity, (e)~experimentally observed transverse intensity profile, (f)~waveguiding in optically induced Bessel photonic lattice. Besides~(c) all figures are normalized.}
\label{fig:bessel}
\end{figure}

Even though these discrete beams seem to be an obvious example for nondiffracting propagation, this fascinating phenomenon historically was discovered by analytical considerations leading to a field distribution radially proportional to a zeroth order Bessel function of the first kind~\cite{ref:Durnin1987}. Up to now, this fundamental Bessel beam is typically generated with a classical optical approach using an axicon~\cite{ref:Scott1992}.

Besides this basic solution of the nondiffracting propagation problem in cylindrical coordinates, field distributions according to higher order Bessel functions~-- described in Fourier space by
\begin{equation}
A(\varphi) = A_0 \exp(\mathrm{i} m \varphi)
\label{eq:fourierspectrum_bessel_general}
\end{equation}
with $m$ being the order~\cite{ref:Bandres2004}~-- show nondiffracting propagation as well~\cite{ref:Bouchal2003}. Only their experimental preparation is much more difficult compared to the zeroth order case.

Nevertheless, our setup provides a simple approach to realize Bessel beams with arbitrary order which in turn are qualified as lattice waves for the generation of photonic structures with circular geometry. According experiments~-- exemplarily shown for a first order Bessel beam ($m = 1$)~-- are summarized in Figure~\ref{fig:bessel}.

Figures~\ref{fig:bessel}a--\ref{fig:bessel}c depict the numerically calculated Fourier spectrum as well as the intensity and phase profiles of the desired nondiffracting beam. The transverse intensity profile of the first order Bessel beam actually generated in our setup and used for the optical induction is shown in Figure~\ref{fig:bessel}e. A subsequent waveguiding experiment (Fig.~\ref{fig:bessel}f) proves the induction of a photorefractive Bessel lattice in a SBN crystal using the nondiffracting light beam (Fig.~\ref{fig:bessel}d). Such an optical potential can for example provide the environment for complex spatial soliton dynamics~\cite{ref:Ruelas2010}.
\section*{Photonic Mathieu lattices}
\label{sec:mathieu}
\begin{figure}[htb]
\centering
\includegraphics{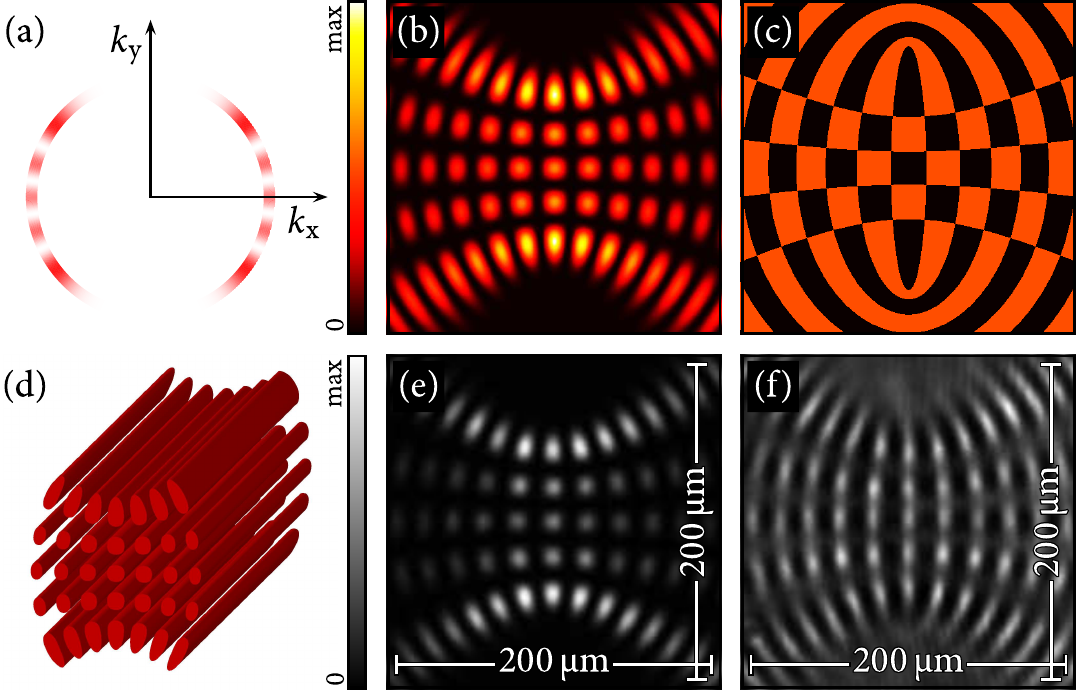}
\caption{Nondiffracting even Mathieu beam ($m = 4$, $q = 6.25 \, \pi^2$, $\varphi_0 = \pi / 2$). (a)~Calculated transverse intensity distribution in Fourier space and (b)~real space, (c)~corresponding real space phase distribution, (d)~iso-intensity surfaces of the nondiffracting light field at 20\% of the maximum intensity, (e)~experimentally observed transverse intensity profile, (f)~waveguiding in optically induced Mathieu photonic lattice. Besides~(c) all figures are normalized.}
\label{fig:mathieu}
\end{figure}

The so-called Mathieu beams constitute the third fundamental family of nondiffracting intensity distributions~\cite{ref:Bouchal2003}. One distinguishes between even and odd Mathieu beams, and their Fourier spectrum is given by
\begin{equation}
A_{\mathrm{e}}(\varphi) = A_0 \, \mathrm{ce}_m(\varphi ,q) \label{eq:fourierspectrum_mathieu_even_general}
\end{equation}
and
\begin{equation}
A_{\mathrm{o}}(\varphi) = A_0 \, \mathrm{se}_m(\varphi, q) \; , \label{eq:fourierspectrum_mathieu_odd_general}
\end{equation}
respectively, where $\mathrm{ce}_m$ and $\mathrm{se}_m$ are the angular Mathieu functions with order $m$ and $q$ is the ellipticity parameter. One member of this complex beam family (an even Mathieu beam with $m = 4$, $q = 6.25 \, \pi^2$ and $\varphi_0 = \pi / 2$) is shown as an example in Figure~\ref{fig:mathieu}.

While Bessel beams are described in cylindrical coordinates, Mathieu beams rely on the elliptic cylindrical coordinate system. This fact becomes manifest in their transverse intensity and phase patterns. Figures~\ref{fig:mathieu}a--\ref{fig:mathieu}c depict these distributions numerically calculated for the selected example. In particular the phase (Fig.~\ref{fig:mathieu}c) illustrates the underlying elliptical character given that the lines separating regions of different phase form ellipses and hyperbolas with joint foci.

Using the numerically calculated intensity and phase distributions as input for our experimental setup, we generate the complex nondiffracting Mathieu beams. For the discussed example, Figure~\ref{fig:mathieu}e shows the corresponding intensity pattern and Figure~\ref{fig:mathieu}f depicts the experimental waveguiding result after optical induction with the distribution in Figure~\ref{fig:mathieu}d.

Besides the manufacturing of complex structures in photosensitive materials, the presented nondiffracting light fields are also outstanding patterning tools. The structured optical potential provided by a complex nondiffracting beam can be used as versatile light moulds for micro particle assemblies as recently shown using Mathieu beams as well~\cite{ref:Alpmann2010}.
\section*{Photonic Weber lattices}
\label{sec:weber}
\begin{figure}[htb]
\centering
\includegraphics{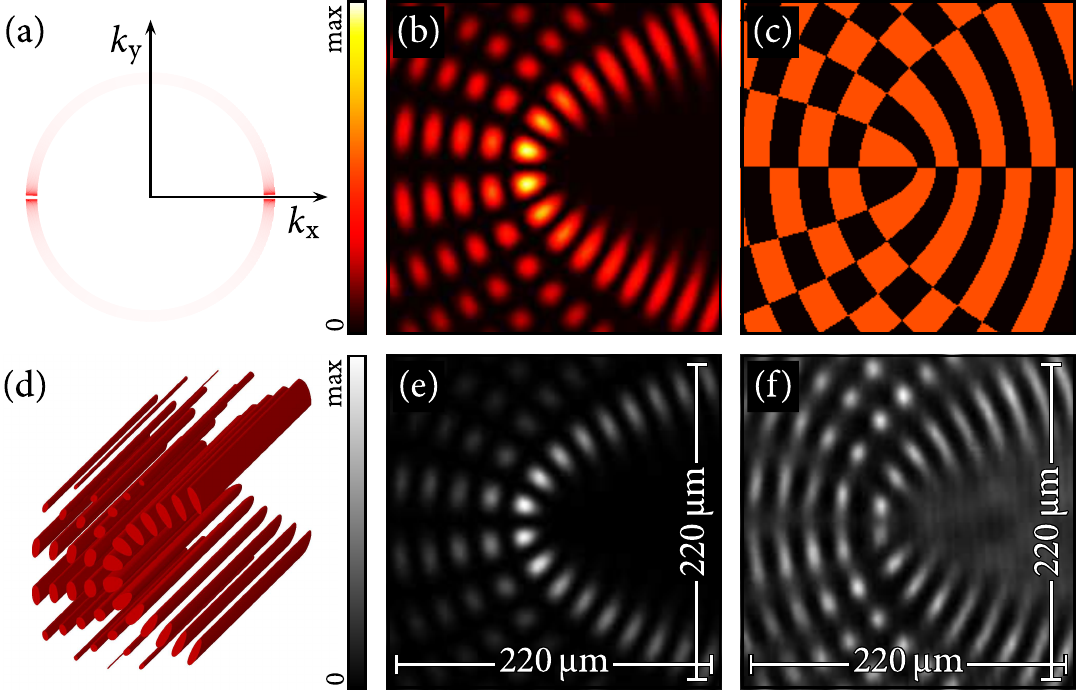}
\caption{Nondiffracting Weber beam ($a = -2$, $\varphi_0 = 0$). (a)~Calculated transverse intensity distribution in Fourier space and (b)~real space, (c)~corresponding real space phase distribution, (d)~iso-intensity surfaces of the nondiffracting light field at 20\% of the maximum intensity, (e)~experimentally observed transverse intensity profile, (f)~waveguiding in optically induced Weber photonic lattice. Besides~(c) all figures are normalized.}
\label{fig:weber}
\end{figure}

The fourth family of nondiffracting intensity distributions aggregates the so-called Weber beams. These beams are based on parabolic cylindrical coordinates and their angular Fourier spectra are given by
\begin{equation}
A_{\mathrm{e}}(\varphi) = \frac{1}{2 \sqrt{\pi \, |\sin\varphi|}} \exp\left(\mathrm{i} a \ln\left| \tan \frac{\varphi}{2} \right| \right) \label{eq:fourierspectrum_weber_even_general}
\end{equation}
and
\begin{equation}
A_\mathrm{o}(\varphi) = \frac{1}{\mathrm{i}}
\begin{cases}
A_{\mathrm{e}} \; , & 0 \leq \varphi \leq \pi\\
-A_{\mathrm{e}} \; , & \pi < \varphi \leq 2 \pi
\end{cases}
\label{eq:fourierspectrum_weber_odd_general}
\end{equation}
for the even and odd case, respectively~\cite{ref:Bandres2004}. The symmetry of a Weber beam depends on the continuous parameter $a$, and Figure~\ref{fig:weber} exemplarily demonstrates a characteristic member of this nondiffracting beam family.

For the experimental realization of this nondiffracting beam, we calculated the required spectrum (Fig.~\ref{fig:weber}a) as well as the intensity (Fig.~\ref{fig:weber}b) and phase (Fig.~\ref{fig:weber}c) profiles and used them again as an input in our setup ($a = -2$, $\varphi_0 = 0$). The actually obtained transverse intensity modulation of the Weber beam is shown in Figure~\ref{fig:weber}e. Finally, the waveguiding image in Figure~\ref{fig:weber}f illustrates that this fourth family of nondiffracting beams is highly applicable to be used for the optical induction of complex two-dimensional photonic lattices inside a photorefractive material as well.

Since the underlying coordinate system~-- just like in the case of the Bessel and Mathieu nondiffracting beams~-- is curvilinear, the induced nonlinear photonic structures share this property and therefore might be an excellent platform for future research on nonlinear photonics in curved systems.

\section*{Conclusions}
\label{sec:conclusions}
In conclusion, we presented a new concept of optical lattice wave generation. For all four families of nondiffracting beams, we are able to transfer calculated field distributions into real nondiffracting lattice waves. Exemplarily, the potential of these complex lattice beams in the field of artificial materials and structures was shown by demonstrating the optical induction of complex discrete, Bessel, Mathieu, and Weber photonic lattices in a nonlinear photorefractive medium.

While all presented photonic structures were designed to be invariant in one distinct direction, our induction concept can easily be modified in order to introduce a modulation in this direction as well~\cite{ref:Xavier2009,ref:Xavier2010}. In addition, techniques known from the field of holographic data storage could be used to multiplex lattices with different feature size, thus facilitating superperiodic structures~\cite{ref:Rose2008} and~-- since the four discussed families of nondiffracting beams each are complete and orthogonal~\cite{ref:Lopez-Mariscal2006}~-- even optical series expansions of arbitrary patterns~\cite{ref:Boguslawski2012}.

Furthermore, the choice of the photosensitive medium is not restricted to the demonstrated photorefractive case. Other materials~-- for example photoresists~\cite{ref:Thiel2010}~-- could also serve as an adequate platform.

At the same time, the presented concept and therewith all the complex nondiffracting beams can be used for the generation of optical potentials in fields like atom optics or cold gases as well. With all this flexibility, our approach has the potential to significantly contribute to modern physics research in multiple fields.
\section*{Acknowledgments}
We acknowledge support by Deutsche Forschungsgemeinschaft and Open Access Publication Fund of University of Münster.
\end{document}